\documentclass[aps,prb,twocolumn,showpacs,amsmath,amssymb,superscriptaddress]{revtex4-1}
\usepackage{graphicx}
\usepackage{dcolumn}
\usepackage{bm}
\usepackage{color}
\usepackage{graphicx}
\usepackage{epsfig}

\newcommand{\bgreek}[1]{\mbox{\boldmath$#1$\unboldmath}}

\setcitestyle{numbers,square}
\include{graphic}

\begin{document}


\title{Coherent control of magnon radiative damping with local photon states}

\author{Bimu Yao}
\affiliation{State Key Laboratory of Infrared Physics, Shanghai Institute of Technical Physics, Chinese Academy of Sciences, Shanghai 200083, People$'$s Republic of China}
\affiliation{Department of Physics and Astronomy, University of Manitoba, Winnipeg, Canada R3T 2N2}

\author{Tao Yu}\email{T.Yu@tudelft.nl}
\affiliation{Kavli Institute of NanoScience, Delft University of Technology, 2628 CJ Delft,  The Netherlands}

\author{Y. S.~Gui}
\affiliation{Department of Physics and Astronomy, University of Manitoba, Winnipeg, Canada R3T 2N2}

\author{J. W.~Rao}
\affiliation{Department of Physics and Astronomy, University of Manitoba, Winnipeg, Canada R3T 2N2}

\author{Y. T.~Zhao}
\affiliation{Department of Physics and Astronomy, University of Manitoba, Winnipeg, Canada R3T 2N2}

\author{W.~Lu}\email{Luwei@mail.sitp.ac.cn}
\affiliation{State Key Laboratory of Infrared Physics, Shanghai Institute of Technical Physics, Chinese Academy of Sciences, Shanghai 200083, People$'$s Republic of China}

\author{C.-M.~Hu}\email{Can-Ming.Hu@umanitoba.ca}
\affiliation{Department of Physics and Astronomy, University of Manitoba, Winnipeg, Canada R3T 2N2}

\date{\today}

\begin{abstract}

The collective excitation of ordered spins, known as spin waves or magnons, can in principle radiate by emitting travelling photons to an open system when decaying to the ground state. However, in contrast to the electric dipoles, magnetic dipoles contributed by magnons are more isolated from electromagnetic environment with negligible radiation in the vacuum, limiting their application in coherent communication by photons. Recently, strong interaction between cavity standing-wave photons and magnons has been reported, indicating the possible manipulation of magnon radiation via tailoring photon states. Here, with loading an yttrium iron garnet sphere in a one-dimensional circular waveguide cavity in the presence of both travelling and standing photon modes, we demonstrate an efficient photon emissions from magnon and a significant magnon radiative damping with radiation rate found to be proportional to the local density of states (LDOS) of photon. By modulating the LDOS including its magnitude and/or polarization, we can flexibly tune the photon emission and magnon radiative damping on demand. Our findings provide a general way in manipulating photon emission from magnon radiation for harnessing energy and angular momentum generation, transfer and storage modulated by magnon in the cavity and waveguide electrodynamics.

\end{abstract}



\maketitle

\section{Introduction}

Magnon is an elementary excitation of magnetic structure that is utilized as information carriers in magnonics and magnon spintronics \cite{magnonics1,magnonics2,magnonics3,magnonics4}, as it carries polarization  or ``spins"  because the magnetization precesses anticlockwise around the equilibrium state \cite{magnonics1,magnonics2,magnonics3,magnonics4}.  Interplay between magnon and other quasiparticles enriches the functionality of information transfer in spintronic devices. Magnons can excite electron spins by the interfacial exchange interaction (spin pumping) \cite{spin_pumping_electron,non_local}, phonons by the magnetostriction \cite{Kamra_phonon,Simon_phonon}, magnons in a proximity magnet through the dipolar or exchange interaction \cite{Yu1,Yu2} and microwave photons by the Zeeman interaction \cite{textbookferrite_JC}. The range of spin-information transfer by quasiparticles is restricted by their coherence length that strongly depends on disorder. Photons are thereby attractive to lift the constraint due to their long coherence time or length in high-quality optical device including the cavity and waveguide. Very recently, pioneering works have combined the best feature of cavity photons and long-lifetime magnon in yttrium iron garnet (YIG) \cite{low_damping,Cao2015}, demonstrating the cavity magnon-polariton dynamics \cite{Soykal2010, Huebl2013, Goryachev2014, Tabuchi2015, zhang2014, Bai2015}. Such high-cooperativity hybrid dynamics stimulates the ideas of coherent information processing with magnon. So far, these works mainly focused on the coherent coupling between magnons and standing-wave photons \cite{Soykal2010, Huebl2013, Goryachev2014, Tabuchi2015, zhang2014, Bai2015,Younonlinear,Yao2017} in confined boundary. However, the efficient delivery of coherent information needs a waveguide \cite{FQ1,FQ2}, in which the magnon radiation at continuous wave range \cite{RD1,RD2,RD3} remains relatively unexplored.

Due to the anguler momentum conservation, the emitted photon by magnon radiation carries the spin current. The accompanying pumping of energy causes the magnon radiative damping \cite{RD1,RD2,RD3}, which reflects the efficiency of such photon emission process \cite{Cont1,Cont2,Cont3}. However, weak coupling between the magnetic dipole and photon leads the radiation of magnon to be relatively difficult \cite{Cont1,Cont2,Cont3}. The magnon-photon interaction is hopefully enhanced by confining photons in a cavity \cite{Soykal2010, Huebl2013, Goryachev2014, Tabuchi2015, zhang2014, Bai2015,Younonlinear, Yao2017} or waveguide. It raises the hope to flexibly control the magnon lifetime on the basis of intrinsic Gilbert damping, helping to overcome the difficulties in controlling the damping and dephasing rates of magnon in conventional solid state with unmanageable elements such as disorder. Moreover, the great tunability of photonic environment in a microwave waveguide could tune the efficiency of pumping the photon spin current from magnon radiation in information processing \cite{FQ1,FQ2}. We envision that in case the mechanism of tuning magnon radiation by local photon states could be demonstrated, various mechanisms that were used to tune photon emission by, for instance, metamaterials, antennas and superconducting circuits could be implemented with magnon to add functionality in magnonic applications \cite{Cont1,MMs,NCS,YouRM}.

In this work, we address a general way to control the photon emission from magnon and magnon radiative damping by tuning the local electromagnetic environment. The radiative damping rate is demonstrated to be proportional to the local density of states (LDOS) of photon in a coupled magnon-photon system. We place an YIG sphere into a circular waveguide cavity that ensembles to a ``clarinet" in shape [see Fig.~\ref{Fig1}(a)]. Similar to the sound physics of ``clarinet",  standing waves are constructed with the superposition of continuous-wave background \cite{Yao2015,APLYAO2015},  highlighting crucial difference with confined cavity in normal coupling scheme. The standing-wave component causes coherent exchange between magnon and photon and induces a splitting gap in the dispersion, while the superposed travelling-wave component play the key role of transferring radiated spin-information to open system.  By simultaneously involving both standing and continuous waves, magnon radiation is thereby effectively controlled by photon states and clearly characterized by magnon linewidth $\Delta H$ from photon transmission. A relative suppression of the radiative damping at cavity resonance is observed that seems to be different from the conventional Purcell effect \cite{Cont3,Cav1,Cav2}, yet is unforeseen in coupled magnon-photon dynamics. These measurements are well explained as we theoretically establish the relation between macroscopic magnon radiative damping and the microscopic LDOS of microwave photons in a quantitative level. Our result opens opportunities to tune the LDOS involving the magnitude and/or polarization to control the photon emission from magnon and magnon radiative damping.  To the best of our knowledge, our work is the first convincing observation of the LDOS-tunable magnon radiative damping in a coupled magnon-photon system, providing the possibility of photon-mediated spin transport with preserved coherence. Due to the linearity nature of our work, we also anticipate that our method offers a general approach to other prototype photonic system or on-chip integrated devices for advancing the manipulation and delivery of radiated spin-information.



\begin{figure*}  [!htbp]
	\begin{center}
		\epsfig{file=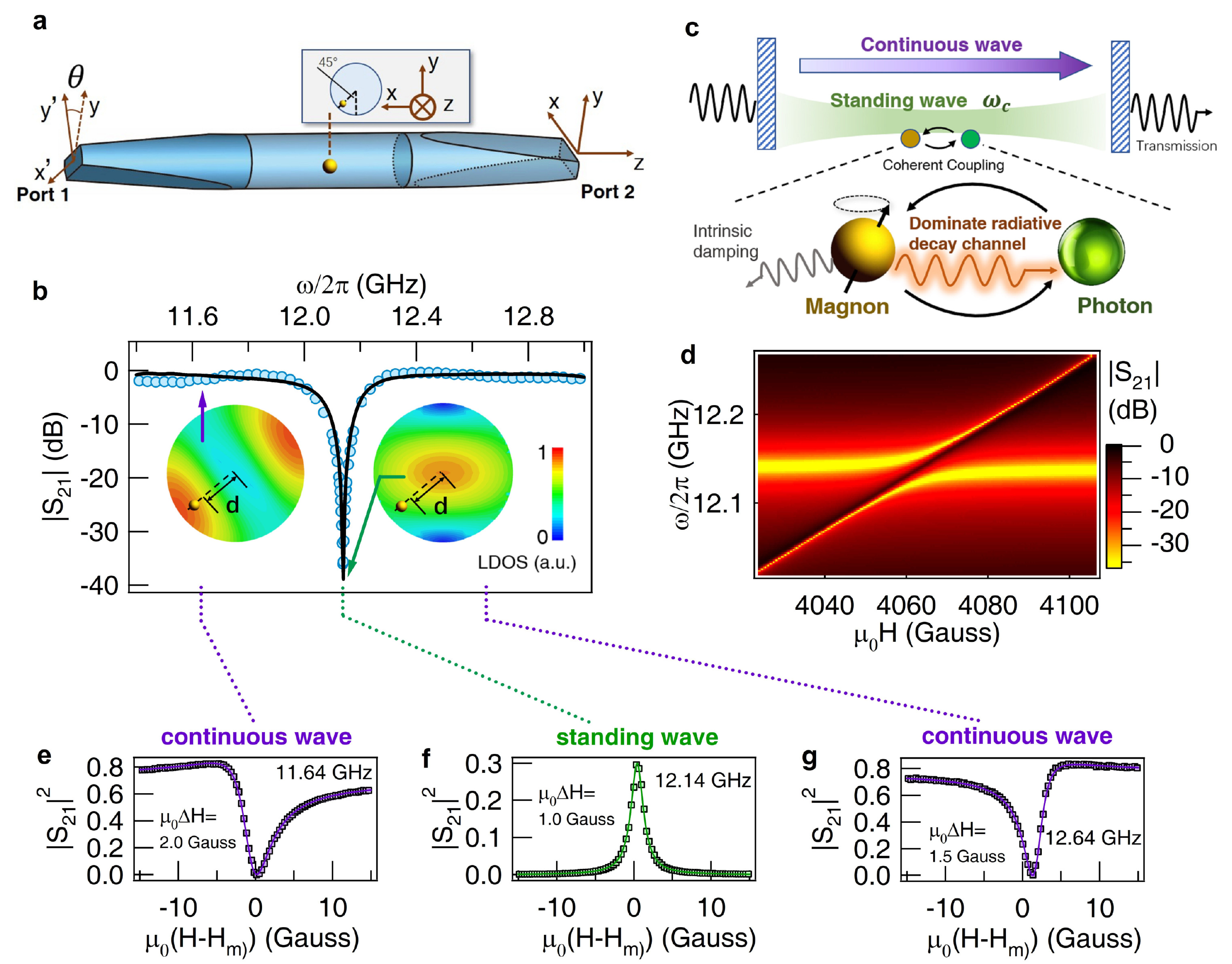,width=16cm}
		\caption{(Color online) \textbf{Magnon radiative damping controlled by LDOS.} (a) Experimental set-up of coupled magnon-photon system in a circular waveguide cavity. (b) Transmission coefficient $|S_{21}|$ from measurement (circles) and simulation (solid lines), with insets showing normalized LDOS distribution for standing-wave resonance 12.14~GHz and continuous wave 11.64~GHz. (c) By coupling magnons with photons in a waveguide cavity, the radiative damping of magnon can be the dominant energy dissipation channel compared to its intrinsic damping. (d) Dispersion map for coupled magnon-photon states. $|S_{21} (H)|^2$ spectra is measured at fixed frequencies 11.64~GHz (e), 12.14~GHz (f) and 12.64~GHz (g), respectively, with the x-axis offset $H_m$ being the biased static magnetic field at magnon resonance. Source data are provided as a Source Data file.}
		\label{Fig1}
	\end{center}
\end{figure*}

\section{Results}
\subsection{Photon states construction}

For clarifying the magnon radiative damping controlled by photon states, we first introduce the local electromagnetic environment inside the circular waveguide cavity as shown in Fig.~\ref{Fig1}(a). This waveguide consists of a 16~mm-diameter circular waveguide and two transitions at both ends that are rotated by an angle of $\theta$=$45^\circ$. The two transitions can smoothly transform TE$_{10}$ mode of  rectangular port to TE$_{11}$ mode of circular waveguide, and vice versa. Specifically, the microwaves polarized in the $\hat{\bf x}$- and $\hat{\bf x}^\prime$-directions are totally reflected at the ends of the circular waveguide, forming the standing waves around specific microwave frequencies. While the microwaves polarized in $\hat{\bf y}$- and $\hat{\bf y}^\prime$-directions can travel across the transitions and therefore form a continuum of travelling waves. Therefore, in our device the standing waves can form around particular wavevectors or frequencies that are superposed on the continuous-wave background \cite{Yao2015,APLYAO2015}. The continuous waves help transfer the information to open system and the standing waves provide the ingredient to form the cavity magnon polariton. Thus, different from discrete modes in the conventional well-confined cavity, our circular waveguide cavity enables to add the ingredient of continuous modes to modify the photonic structure \cite{Yao2015}.

The modes in our device can be characterized by microwave transmission using a vector network analyzer (VNA) between port 1 and 2. A standing wave or ``cavity" resonance mode at $\omega_c/2\pi$=12.14~GHz is clearly revealed in $S_{21}$ with a loaded damping factor of $9\times10^{-3}$, as illustrated by blue circles in Fig.~\ref{Fig1}(b). This damping of photon (109.3~MHz) is larger than the coupling strength between standing-wave photon and magnon, and thus our system lies in the magnetically induced transparency (MIT) rather than the strong regime \cite{zhang2014}. Accordingly, even the standing wave allows the delivery of energy out of the waveguide through its damping. It is observed in the transmission spectrum that standing waves confined in the waveguide cause the dip in transmission spectrum at cavity resonance \cite{Yao2015}, while the travelling continuous waves that deliver photons from port 1 to 2 contribute a high transmission close to 1. Since continuous waves are not negligible in our device, photon modes thereby cannot be described by a single harmonic oscillator as shown in previous works\cite{Huebl2013, Goryachev2014, Tabuchi2015, zhang2014, Bai2015}. Hence the electromagnetic fields in our waveguide cavity are described by a large number of harmonic modes \cite{M1,M2,M3} in a wide frequency range and each mode has a certain coupling strength with magnon mode.

The following Fano-Anderson Hamiltonian describes the interaction between magnon and photon as \cite{M1,textbookferrite_JC}
\begin{equation}\label{ham}
H_0/\hbar=\omega_m \hat{m}^\dag \hat{m}+\sum_{k_z}\omega_{k_z}\hat{a}_{k_z}^{\dagger}\hat{a}_{k_z}+\sum_{k_z}g_{k_z} (\hat{m}^\dag\hat{a}_{k_z}+\hat{m}\hat{a}_{k_z}^{\dagger}),
\end{equation}
where $\hat{m}^\dag$ ($\hat{m}$) is the creation (annihilation) operator for magnon in Kittel mode with frequency $\omega_m$, $\hat{a}_{k_z}^{\dagger}$ ($\hat{a}_{k_z}$) denotes the photon operator with wavevector $k_z$ and frequency $\omega_{k_z}$, and $g_{k_z}$ represents the corresponding coupling strength between the magnon and microwave photon mode.  We visualize magnon Kittel mode as a single harmonic oscillator in Eq.~(\ref{ham}). Magnon and photon modes have intrinsic dampings originated from inherent property, but our cavity establishes coherent coupling between them \cite{RD1,RD2,RD3} as schematically shown in Fig.~\ref{Fig1}(c).

Due to the coherent coupling between magnon and photon, the energy of excited magnon would radiate to the photons that travel away from the magnetic sphere. This can be pictured as the ``auto-ionization" of magnon into the propagating continuous state that induces the photon emission from magnon and hence magnon radiative damping \cite{AT1,AT2}. Such ``additional" magnon dissipation induced by photon states can be rigorously calculated by the imaginary part of self-energy in magnon Green's function that is expressed as $\Delta E_m=\delta _m+\frac{\pi}{\hbar}|\hbar g(\omega)|^2D(\omega)$. Here, $\delta _m$ is the intrinsic dissipation rate of magnon mode, $D(\omega)$ represents the global density of states for the whole cavity that is a count of the number of modes per frequency interval. We note that above radiative damping is established when the on-shell approximation is valid with energy shift of magnon (tens to hundreds of MHz) is much smaller than its frequency (several GHz) (see Supplementary Note 1) \cite{AT1,AT2}. By further defining the magnon broadening in terms of magnetic field $\Delta E=\hbar\gamma \mu_0 \Delta H$, the magnon linewidth is expressed as (Supplementary Note~1)
\begin{equation}
\label{main}
\mu_0\Delta H=\mu_0\Delta H_0+\frac{\alpha\omega}{\gamma}+\frac{2\pi \kappa}{\gamma}R |\rho_l(d,\omega)|,
\end{equation}
in which $\gamma$ is modulus of the gyromagnetic ratio and $\mu_0$ denotes the vacuum permeability.
In Eq.~(\ref{main}), the first two terms are the linewidth related to inherent damping of magnon in which $\mu_0\Delta H_0$ and $\alpha\omega/\gamma$ come from the inhomogeneous broadening
at zero frequency \cite{H0} and the intrinsic Gilbert damping, respectively. The last term describes the radiative damping induced by photon states in which $|\rho_l(d,\omega)|$ represents the LDOS of magnetic fields with $d$ and $l$ denoting the position and photon polarization direction.  Basically, $|\rho_l(d,\omega)|$ counts both the local magnetic field strength and the number of electromagnetic modes per unit frequency and per unit volume. $\kappa=\frac{\gamma M_sV_s}{2\hbar c^2}$ with $M_s$ and $V_s$ being the saturated magnetization and volume of the loaded YIG sphere. $R$ represents a fitting parameter that is mainly influenced by cavity design and cable loss in the measurement circuit.

Based on above theoretical analysis, we find that the radiative damping is exactly proportional to the LDOS $\rho_l(d,\omega)$. To observe radiation as a dominant channel for the transfer of magnon angular momentum, it requires both low inherent damping of magnon and a large tunable $|\rho_l(d,\omega)|$. In the following experiment, both conditions are satisfied by introducing a YIG sphere with low Gilbert damping, as well as by modifying photon mode density through tuning LDOS magnitude (Sec.~\ref{magnitude}), LDOS polarization (Sec.~\ref{polarization}) and global cavity geometry (Sec.~\ref{global}).

\subsection{Magnon radiation tunned by photon LDOS}

A highly polished YIG sphere with 1~mm diameter is loaded into the middle plane of waveguide cavity. Before immersing into experimental observations, it is instructive to understand the two-dimensional (2D) spatial distribution of LDOS in the middle plane, which is numerically simulated by CST (\emph{Computer simulation technology}) at the center cross section that can well-reproduce $|S_{21}|$ as shown in Fig.~\ref{Fig1}(b). It can be seen that  the hot spots for the continuous waves (11.64~GHz) and standing wave (12.14~GHz) are spatially separated,
 providing the possibility to control LDOS magnitude by tuning the positions of magnetic sample inside the cavity.

In our first configuration, we focus on the local position with $d$=6.5~mm as marked in Fig.~\ref{Fig1}(b). Such position enables the magnon mode not only to have overlapping \cite{zhang2014} with standing waves but also to couple to the continuous ones. More interestingly, as indicated by the insets in Fig.~\ref{Fig1}(b), LDOS at $d$=6.5~mm has smaller quantity at cavity resonance compared with the ones in the continuous-wave range, which is opposite to the LDOS enhancement at resonance in conventional well-confined cavity \cite{Cont3,Cav1,Cav2}. Therefore, according to Eq.~(\ref{main}), in contrast to magnon linewidth enhancement in previous works, we expect a totally different linewidth evolution by varying frequency in measurement with linewidth suppression at cavity resonance $\omega_c$. 

Concretely, the magnon linewidth can be measured from $|S_{21}|$ spectra in $\omega$-$H$ dispersion map. In our measurement, a static magnetic field $\mu_0H$ is applied along the $\hat{\bf x}$-direction to tune the magnon mode frequency (close to or away from the cavity resonance), which follows a linear dispersion $\omega_m=\gamma\mu_0(H+H_A)$ with $\gamma=2\pi\times28$ GHz/T and $\mu_0H_A$=192~Gauss being the specific anisotropy field. For our YIG sphere, the saturated magnetization is $\mu_0 M_s$=0.175~T, the Gilbert damping $\alpha$ is measured to be $4.3\times10^{-5}$ by standard waveguide transmission with the fitted inhomogeneous broadening $\mu_0\Delta H_0$ being 0.19 Gauss. As $\omega_m$ is tuned to approach the cavity resonance $\omega_c$, a hybrid state is generated with the typical anti-crossing dispersion as displayed in Fig.~\ref{Fig1}(d). A coupling strength of 16~MHz can be found from rabi splitting at zero detuning condition that indicates the coherent energy conversion between magnon and photon.

\begin{figure}  [b]
	\begin{center}
		\epsfig{file=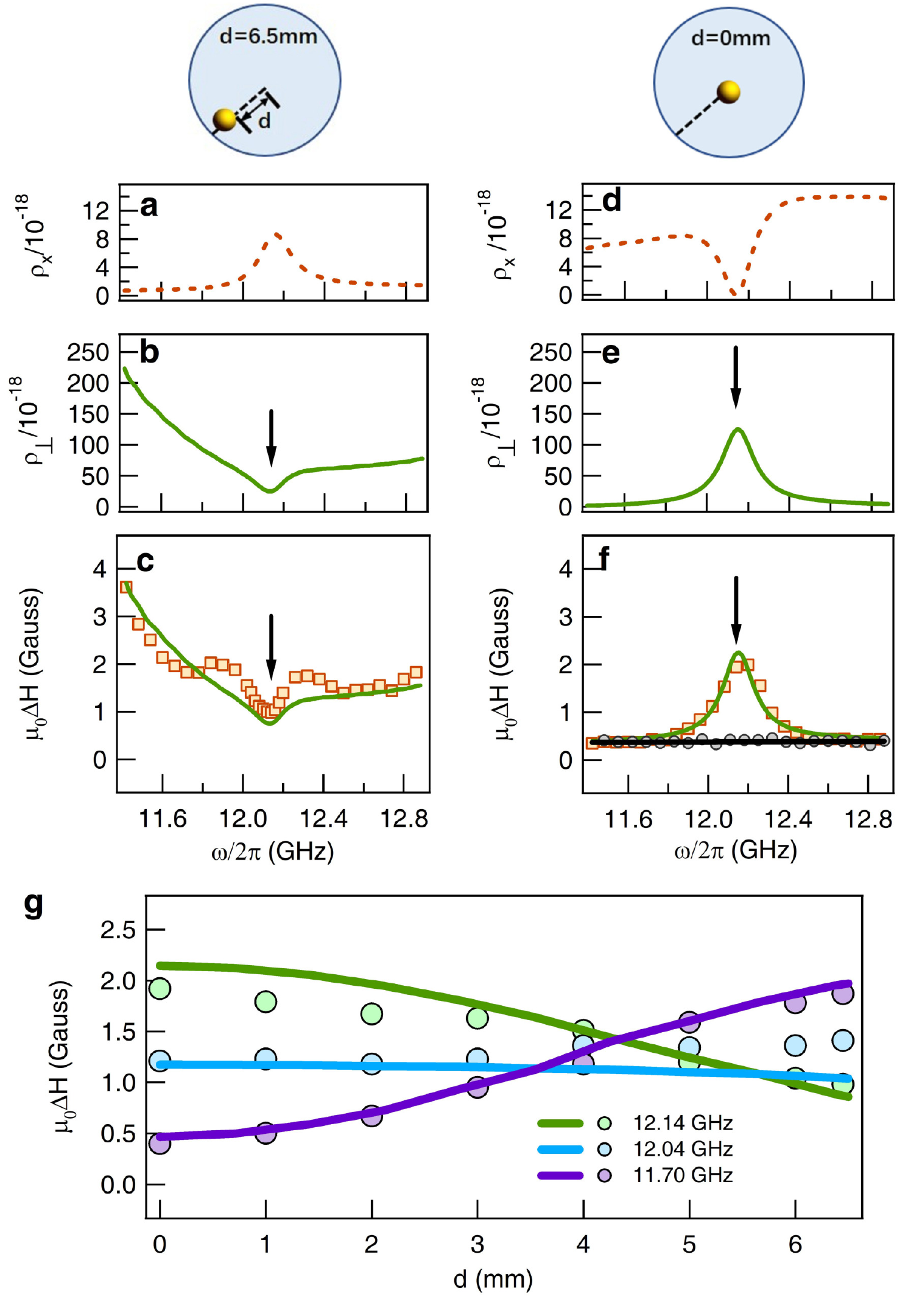,width=8.7cm}
		\caption{(Color online) \textbf{LDOS magnitude dependence.} (a) and (b), Simulated $\rho_x$  and $\rho_\perp$ at $d$=6.5~mm. (c) Measured $\mu_0\Delta H$-$\omega$ relation (squares) with calculated lines from model (the green line) at $d$=6.5~mm.  (d) and (e), Simulated $\rho_x$ and $\rho_\perp$ at $d$=0~mm. (f) Measured $\mu_0\Delta H$-$\omega$ relation (squares) with calculated lines from model (the green line) at $d$=0~mm. Black circles and lines indicate the measured and fitted intrinsic linewidth, respectively. (g) $\mu_0\Delta H$ evolution with tuning positions for different frequencies, with circles and solid lines representing the measured magnon linewidth and the linewidth computed from LDOS, respectively. Source data are provided as a Source Data file.}\label{Fig2}
	\end{center}
\end{figure} 

Magnon linewidth (HWHM) is characterized by a lineshape fitting of $|S_{21}(H)|^2$ that is obtained from the measured transmission at fixed frequency and different magnetic fields. Here, we focus on $|S_{21}(H)|^2$ at three different frequencies with one being at the cavity resonance $\omega_c$ and the other two chosen at continuous wave frequencies above and below $\omega_c$ (11.64~GHz and 12.64~GHz, respectively). As photon frequency is tuned from continuous-wave range to the cavity resonance $\omega_c/2\pi$=12.14 GHz, we observe that the lineshape of $|S_{21}(H)|^2$ varies from asymmetry to symmetry, accompanied by an obvious linewidth suppression from 2.0~Gauss/1.5~Gauss to 1.0 Gauss as shown in Fig.~\ref{Fig1}(e)-(g).

It is worth noticing that the magnon linewidth $\mu_0\Delta H$ shows a clear suppression at cavity resonance rather than the linewidth enhancement in conventional coupled magnon-photon system in the cavity \cite{Bai2015, BaiIEEE}. Such suppression of magnon linewidth qualitatively follows the LDOS magnitude, which also shows smaller quantity at cavity resonance. This qualitatively agrees with our theoretical expectation from Eq.~(\ref{main}). In the following subsections, it is necessary to study the relation between linewidth and LDOS in a quantitative level by using both theoretical calculation and experimental verification.

\begin{figure*}  [!htbp]
	\begin{center}
		\epsfig{file=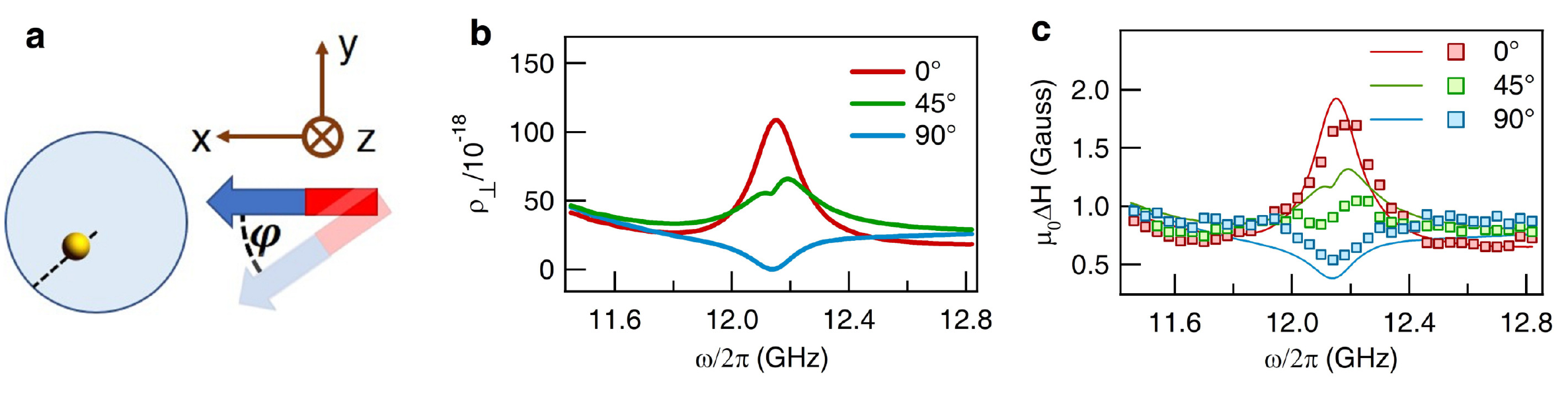,width=16 cm}
		\caption{(Color online) \textbf{LDOS polarization dependence.} (a) Schematic of tuning orientation of external magnetic field $H$ relative to the $\hat{\bf x}$-direction in the plane of waveguide cross section (b) Simulated photon LDOS perpendicular to external magnetic field $H$ with relative angle $\varphi=0^\circ$, $45^\circ$ and $90^\circ$, respectively. (c) Measured magnon linewidth spectra, i.e., $\mu_0\Delta H$-$\omega$ relation (squares) and calculated results (solid lines) for different angles $\varphi=0^\circ$, $45^\circ$ and $90^\circ$, respectively. Source data are provided as a Source Data file.}\label{Fig3}
	\end{center}
\end{figure*}

\subsubsection{Magnon radiation controlled by LDOS magnitude}
\label{magnitude}

In this subsection, we show a quantitative control of magnon radiative damping by tuning the LDOS magnitude over a broadband frequency range. The spacial variation of the magnetic field in our waveguide cavity allows us to realize different LDOS spectra by simply choosing different positions. Similar to the experimental settings in the above section with $d$=6.5~mm, we display a broadband view of LDOS for per polarization by using simulation in Fig.~\ref{Fig2}. Although $\rho_x(\omega)$   in Fig.~\ref{Fig2}(a) shows a typical resonance behaviour, its contribution to the magnon radiation is negligible here according to the well-known fact that only photon polarization that perpendicular to the external static magnetic field $H$ drives the magnon dynamics. Following this consideration, we further simulate $\rho_\perp$=$\sqrt{\rho_y^2+\rho_z^2}$ that plays a dominant important role in the magnon-photon interaction as displayed in Fig.~\ref{Fig2}(b). $\rho_\perp(\omega)$ shows a dip at the cavity resonance in the frequency dependence.

It is clearly seen that due to the enhancement of global density of states at the mode cut-off of the waveguide, continuous wave LDOS becomes more and more significant when frequency goes lower to approach the cut-off frequency (around 9.5~GHz). This phenomenon can be viewed as a Van Hove singularity effect in the density of states for photons (see independent observation via a standard rectangular waveguide in Supplementary Note~2). As such singularity effect is involved in the coupled magnon-photon dynamics, we obtain larger linewidth at detuned frequency range that causes a clear linewidth suppression at cavity resonance. In sharp contrast to linewidth enhancement from typical Purcell effects, results in Fig.~\ref{Fig2}(c) provides a new linewidth evolution process in a broadband range. Furthermore, to compare with our theoretical model, we perform calculation by Eq.~(\ref{main}) with $\kappa R=4.0\times 10^{22}$~$\rm m^3/s^{2}$ by using the fitting parameter quantity $R\sim 0.8$. It can be observed in Fig.~\ref{Fig2}(c) that the measured $\mu_0\Delta H$ agrees well with the computed ones from our theoretical model. The reasonable agreement between experiment and theory obtained with $R$ close to unity suggests that the linewidth is coherently controlled by LDOS magnitude, especially showing that radiative power emission induced by continuous waves can unambiguously exceed that induced by standing waves.


To create a different LDOS magnitude to tune the magnon radiation, the magnetic sphere is moved to the center of the cross section with $d$=0~mm. The simulated LDOS $\rho_x$ and $\rho_\perp$ are illustrated in Fig.~\ref{Fig2}(d) and (e), respectively. The effective LDOS $\rho_\perp$ shows an enhancement at cavity resonance but suppressions at continuous-wave range.
Similar to the frequency dependence of the LDOS magnitude, the magnon linewidth is clearly observed  to be enhanced at cavity resonance but suppressed at continuous waves. This relation between the magnon width and LDOS is again quantitatively verified by the good agreement between measurement and calculated results from Eq.~(\ref{main}) as shown in Fig.~\ref{Fig2}(f). Particularly, as continuous wave LDOS is suppressed to nearly zero, the radiative damping from LDOS thereby becomes negligibly small. In this case, it can be found that the magnon linewdith exactly returns to its intrinsic damping $\mu_0\Delta H_0+\alpha \omega/\gamma$ mearsured in an independent standard waveguide. 

Finally, at a detailed level, to continuously tune the ratio of standing/continuous-wave LDOS magnitude, the position of the YIG sphere is moved with $d$ varied from 0~mm to 6.5~mm. Typically for three different frequency detunings with 0~MHz, -100~MHz and -440~MHz, our results in Fig.~\ref{Fig2}(g) shows that the magnon linewidth can be controlled with enhancement, suppression or negligible variation in the position dependence. As shown in Fig.~\ref{Fig2}(g), these results showing good agreement with the theoretical calculation suggests that magnon linewdith can be controlled on demand by tuning the LDOS magnitude. Moreover, the photon emission efficiency from magnon radiation can in principle be significantly enhanced with a larger magnetic sphere and smaller waveguide in cross section. For example, a magnetic sphere in 2-mm diameter and a waveguide with half radius would enhance the radiation rate by 16 times (Supplementary Note 1).

\subsubsection{Magnon radiation controlled by LDOS Polarization}
\label{polarization}

Having shown the relation between the magnon radiative damping in $\mu_0\Delta H$ and the LDOS magnitude, here we would like to introduce LDOS polarization as a new degree of freedom to control the magnon radiation. In our experiment, by placing the YIG sphere at $d$=2.3~mm, the tuning of effective LDOS polarization $\rho_\perp$ for the magnon to feel can be simply achieved by varying the direction of external static magnetic field $H$ with a relative angle $\varphi$ to the $\hat{\bf x}$-direction as shown in Fig.~\ref{Fig3}(a). Please note that compared with the complicated operation to vary the position of YIG sphere inside a cavity, here the LDOS was controlled continuously in a large range simply by rotating the orientation of the static magnetic field. Based on the orthogonal decomposition of LDOS for photons,  $\rho_\perp$ is simulated for three typical angles $\varphi$, i.e., 0$^\circ$, 45$^\circ$, and 90$^\circ$ as shown in Fig.~\ref{Fig3}(b).  For $\varphi=0^\circ$ with $H$ being exactly in the $\hat{\bf x}$-direction, the LDOS is dominated by standing-wave component, which could provide largest coupling with magnon at cavity resonance. While as $\varphi$ goes close to 90$^\circ$, continuous waves become more and more dominant in the contribution to the LDOS, causing a peak-to-dip flip for LDOS around $\omega_c$ in Fig.~\ref{Fig3}(b).

Accordingly, in our experiment, we obtain a magnon linewidth enhancement at $\varphi=0^\circ$ as shown in Fig.~\ref{Fig3}(c) by red squares. As $\varphi$ is tuned towards 90$^\circ$, we thereby anticipate and indeed obtain a linewdith suppression at cavity resonance with blue squares, showing good agreement with the linewidth scaling of $\rho_\perp$ in Eq~(\ref{main}). The theoretically calculated linewidth $\mu_0\Delta H$ is plotted for each $\varphi$ in Fig.~\ref{Fig3}(c) with $\kappa R$ being consistent with the previous subsection. The good agreement between experimental and theoretical findings suggests a flexible control of magnon radiation via LDOS polarization. Moreover, not restricted to tune relative angle between $H$ and LDOS polarization in 2D plane, more possibility of magnon radiation engineering may be realized by pointing $H$ to arbitrary direction in the whole 3D space.

\begin{figure}  [t]
	\begin{center}
		\epsfig{file=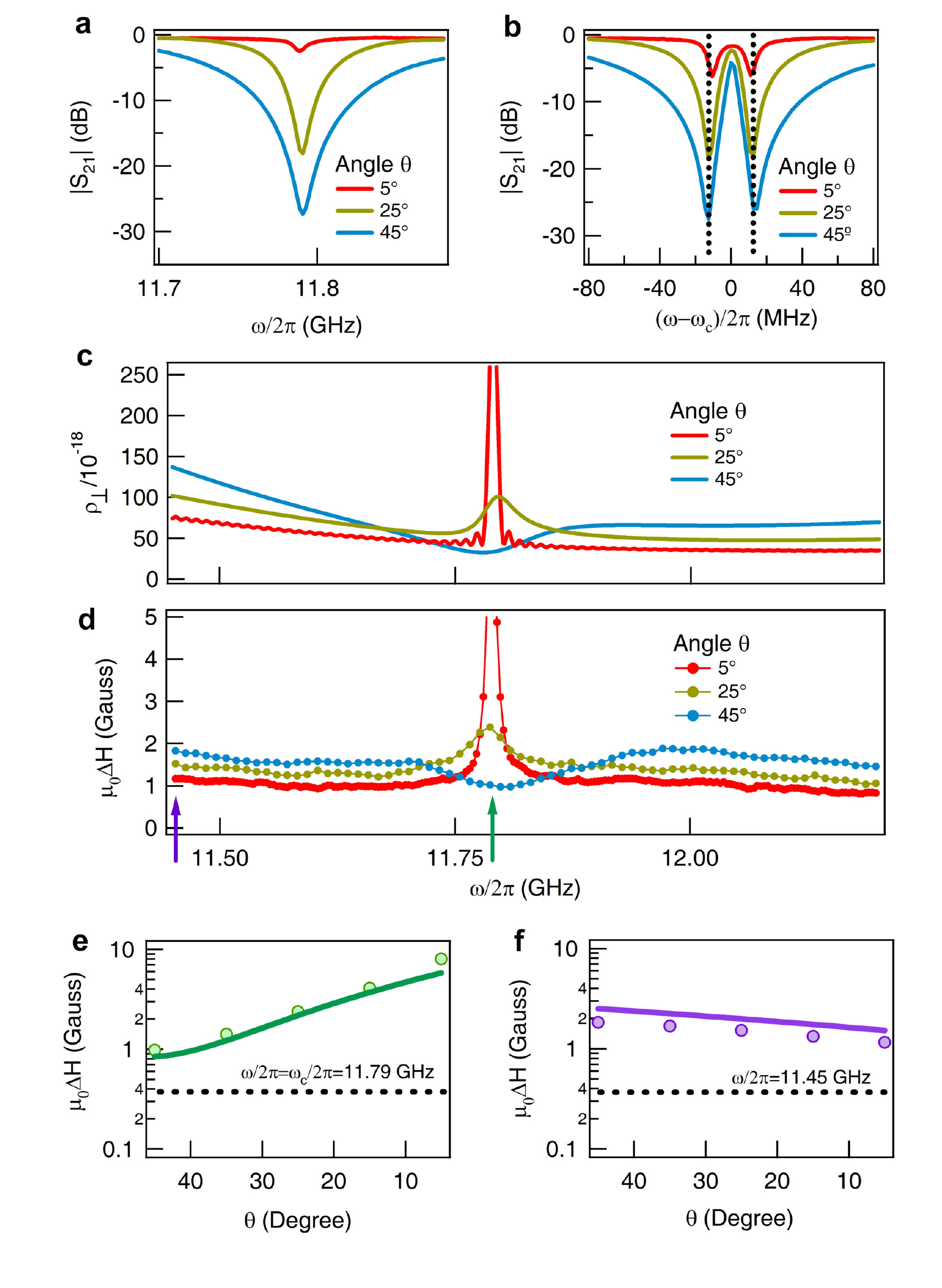,width=8.7 cm}
		\caption{(Color online) \textbf{Cavity geometry dependence.} (a) Cavity mode transmission profile when rotating $\theta$. (b) Rabi splitting spectra for different angles $\theta$.  (c) Simulated $\rho_\perp$ for different $\theta$. (d) Measured $\mu_0\Delta H$-$\omega$ relation when tuning $\theta$.  (e) and (f) Comparison between  theoretical results and measurement at cavity resonance 11.79~GHz (e) and continuous wave frequency 11.45~GHz (f). Dashed lines are intrinsic linewidth of YIG sphere. Source data are provided as a Source Data file.}
		\label{Fig4}
	\end{center}
\end{figure}

\subsubsection{Magnon radiation controlled by cavity geometry}
\label{global}

Our device allows us to tune the LDOS magnitude and polarization together by simply rotating the relative angle $\theta$ between the two transitions \cite{Yao2015}, i.e., the global geometry of our circular waveguide cavity. This can again validate and enrich our observations that the same magnon harmonic mode radiates a different amount of power depending on the surrounding photon environments. In this subsection, we insert a rotating part in the middle plane of cavity, so that the relative angle $\theta$ between two transitions can be smoothly adjusted. By tuning angle $\theta$ from 45 degree to 5 degree, our system shows a significant change in photon transmission as illustrated in Fig.~\ref{Fig4}(a), accompanied by significant enhancements in cavity quality factor and global density of states \cite{Wigner, Smith}. In addition, cavity resonance shows red shift to 11.79~GHz due to the increase of cavity length.
The YIG sphere is placed at cavity center cross section with $d$=6~mm and the external magnetic field is applied in the $\hat{\bf x}$-direction. Such experimental conditions provide stable magnon-photon coupling strength when $\theta$ is tuned, as shown by the nearly unchanged mode splitting in Fig.~\ref{Fig4}(b).

Our hybrid system now readily allows us to the investigate the magnon radiation controlled by cavity geometry. In particular, tuning $\theta$ from 45 degree to 5 degree leads to a redistribution of photon states in cavity that greatly enhances the LDOS near $\omega$=$\omega_c$ and controls the continuous wave LDOS in an opposite way, as illustrated by the simulated $\rho_\perp$ in Fig.~\ref{Fig4}(c). Based on the theoretical model, we expect magnon linewidth can quantitatively follow the geometry-controlled $\rho_\perp$.  Results from measurements under different $\theta$ are shown in Fig.~\ref{Fig4}(d) and we indeed obtain linewidth $\mu_0\Delta H$ with similar behavior to the simulated $\rho_\perp$. As is evident in Fig.~\ref{Fig4}(e) and (f), we find that the linewidth is well reproduced by our theoretical model with $\kappa R$ adjusted to $4.3\times 10^{22}$~$\rm m^3/s^{2}$. By tuning LDOS via $\theta$, the experimental linewidth is enhanced by twenty-fold at cavity resonance in comparison with the intrinsic damping of magnon as illustrated by the dashed lines.

\section{Discussion}

Understanding and controlling magnon radiative damping is essential in tuning the magnon lifetime as well as transporting spin information by travelling photon in spintronic or magnonic applications \cite{Cont1,Cont2,Cont3,SLM1,SLM2}. With revealing the quantitative relation between magnon radiative damping and photon LDOS for the first time, our work brings three perspectives for better exploring and utilizing the magnon radiation in future research.

(i) \textbf{Flexible control of magnon lifetime.} Photon LDOS construction can flexibly tune the magnon lifetime on the basis of intrinsic Gilbert damping. Although long magnon lifetime is useful for information storage and memory, suppressed magnon lifetime would bring advantageous impact for realizing fast repetition rate in device \cite{Cont3}. Our work explores some techniques including LDOS magnitude, polarization and global environment to control magnon lifetime, which could open paths for various LDOS control method to tune magnon radiation in a flexible and precise fashion, and provids a new ingredient to advanced communication processing \cite{Darkmem}.

(ii) \textbf{Delivering coherent information of cavity-magnon polariton to open system.} Although the weak interaction between magnetic dipole and photons can be enhanced by confining the photon mode in a cavity, the confinement restricts the magnon radiated information to be efficiently transferred to the open system and vice versa. Our constructed magnon-photon system can combine both standing and traveling photon modes to couple the magnon. The traveling channel allows to deliver the coherent information out and advances the efficient tuning on the dynamics of the cavity-magnon polariton. Manipulation of magnon radiation to open system, as we demonstrated, is very attractive to explore new physics related to magnon dissipative procession \cite{MH18}, such as dissipative coupling in magnon-based hybrid system \cite{APEnaka}.

(iii) \textbf{Stimulating the advancement of hybrid magnonics.} Controllable magnon radiation could stimulate hybrid magnonic systems to access new frontiers. Recent study shows coherent information at single magnon level can be coherently transferred to photon or superconducting qubit through radiation at millikelvin temperatures \cite{Tabuchi2015,ta2014}, bringing quantum nature to the hybridized magnonic system. At room temperature, generally the magnon-photon coupling is restricted in linear harmonic dynamics, while our recent research breaks this harmonic restriction by using feedback mechanism, exhibiting nonlinear triplet spectra similar to quantum dots \cite{Yao2017}. The magnon radiation in these new regimes could stimulate the advancement of hybrid magnonics.

In conclusion, we observe and show the ability to control photon emission from magnon and magnon radiative damping in the hybrid magnon-photon system, bridging their relation to the tunable photon LDOS.  Compared with conventional enhancement of magnon damping at cavity resonance in well-confined magnon-photon system, we report that the magnon linewidth at cavity resonance can be relatively suppressed by photon LDOS engineering. One quantitative method to design and tune the radiation efficiency of magnon is provided based on tailoring photon LDOS including LDOS magnitude and/or polarization, thereby leading to a general technique of tuning magnon relaxation on demand. Our measurements are mainly performed in the MIT regime with large photon damping, which causes the radiative damping by photon dissipation; while travelling-wave photon can directly transfer the magnon energy to open system. Overall, our study introduces a novel mechanism to coherently manipulate magnon dynamics by local photon states and suggests a promising potential towards the development of magnon-based hybrid devices and related coherent information processing.

\section{Methods}

\subsection{Device description}
Our waveguide cavity is made up of a cylindrical waveguide and two circular rectangular transitions coaxially connected at both ends. Through the transition, a smooth change between TE$_{10}$ mode of rectangular waveguide port and TE$_{11}$ one of cylindrical waveguide can be established. Via coaxial cables, the cavity is connected to the input/output ports of an vector network analyzer (VNA). With an input power of 0~dBm, the transmission signals can be precisely picked up by VNA. YIG sphere is fixed firmly inside the cavity with scotch tape, with its location tunable on demand to couple with different microwave magnetic fields. YIG and the scotch tape, as dielectric materials, can slightly influence the microwave fields distribution in our experiment. We neglect the small dielectric influence in our theoretical treatment.


\subsection{Theoretical description}
In Supplementary Note 1, the theory of magnon spontaneous radiation in the waveguide including the derivation of the magnon linewidth induced by the LDOS is provided.

\section{Data availability}
The source data underlying Figs. 1-4 and Supplementary Figs. 1-2 are provided as a Source Data file. The data that support the findings of this study are available from the corresponding authors upon reasonable request.

\section{Acknowledgement}
This work was funded by NSERC, the National Natural Science Foundation of China under Grant No.11429401 and No.11804352, the Shanghai Pujiang Program No.18PJ1410600, the Science and Technology Commission of Shanghai Municipality (STCSM No. 16ZR1445400), and  SITP Innovation Foundation (CX-245). T.Y. was supported by the Netherland Organization for Scientific Research (NWO). We would like to thank Y. Zhao, J. Sirker, L. H. Bai, P. Hyde, Y. M. Blanter and G. E. W. Bauer for useful discussions.

\section{Author contributions}
B.M.Y. and Y.S.G. set up the hybrid system, conducted the experiment as well as analyzed the data. T.Y., in discussions with B.M.Y. and C.-M.H., developed the theory part.  J.W.R. and Y.T.Z. contributed to the design of the cavity parts.  B.M.Y. and T.Y. prepared all the figures as well as the supplementary material part. T.Y., L.W. and C.-M.H. together supervised the work. All authors contribute to the paper writing. 

\begin{widetext}
\section{Supplementary Note 1.\\ Theory of magnon spontaneous radiation}

In this part, we describe the model about the magnon spontaneous radiation in a waveguide. We first set up the Fano-Anderson Hamiltonian for the magnon-photon coupling in a waveguide \cite{Fano,Mahan}, and then calculate the energy broadening due to the magnon spontaneous radiation.

\subsection{Hamiltonian}

The free energy of the coupled system is written as \cite{electromagnetism}
\begin{equation}
F=F_0({\bf M})+\frac{\epsilon_0}{2}\int d{\bf r}{\bf E}({\bf r})\cdot{\bf E}({\bf r})+\frac{\mu_0}{2}\int d{\bf r}{\bf H}({\bf r})\cdot{\bf H}({\bf r})+{\mu_0}\int d{\bf r}{\bf H}({\bf r})\cdot{\bf M}({\bf r}),
\end{equation}
where $\epsilon_0$ and $\mu_{0}$ are the vacuum
dielectric and permeability constants, respectively. $F_0({\bf M})$ describes the free energy of magnetization ${\bf M}$, the second and third terms denote the free energy of electromagnetic waves due to the electric field ${\bf E}$ and magnetic field ${\bf H}$ in a waveguide, and the last term represents the Zeeman interaction between the magnetization and the magnetic field. The Hamiltonian can be obtained by quantizing the free energy.

Due to the long-wavelength nature of the electromagnetic waves in the waveguide we are considering, only the uniform precession of the magnetization, i.e., the Kittel mode \cite{Kittel_book}, needs to be considered. We express
${\bf M}=-\gamma\hbar {\bf S}$ in terms of the spin operators \cite{Kittel_book,squeezed_magnon,HP}, where $\gamma$ is modulus of the gyromagnetic ratio. The spin operators are expanded by magnon operator $\hat{\alpha}_K$
in the Kittel mode \cite{Kittel_book,squeezed_magnon,HP},
\begin{equation}
\hat{S}_{\beta}^K=\sqrt{2S}\big[M_{\beta}^K\hat{\alpha}_K(t)+(M_{\beta}^K)^*\hat{\alpha}^{\dagger}_K(t)\big],
\end{equation}
in which $\beta=\{z,x\}$ assuming that the equilibrium magnetization is along the $\hat{\bf y}$-direction.
$S=M_s/(\gamma \hbar)$ with $M_s$ being the saturated magnetization;  $M_{\beta}^K$ is the amplitude of the Kittel mode that is normalized according to \cite{Walker,magnetic_nanodots}
\begin{equation}
\int d{\bf r}\Big[M_z^K({\bf r})(M_x^K({\bf r}))^*-(M_z^K({\bf r}))^*M_x^K({\bf r})\Big]=-i/2.
\end{equation}
Because the magnetization is uniform in the magnetic sphere and the Kittel mode is circularly polarized with $M_x=iM_z$, we obtain
\begin{equation}
M_z^K=1/(2\sqrt{V_s}),
\end{equation}
in which $V_s$ is the volume of the magnetic sphere.

In the waveguide, the electromagnetic field operators are expanded by the photon operator $a_{k,\lambda}$ in which $k>0$ is the momentum of photon along the waveguide, i.e., the $\hat{\bf z}$-direction,
and $\lambda$ ($\mu$ below) labels the eigenmodes in the waveguide \cite{electromagnetism},
\begin{eqnarray}
\nonumber
\hspace{-0.6cm}&&{\bf E}({\bf r})=\sum_{\lambda,k}\big[{\bf E}_k^{\lambda}(x,y)e^{ikz}\hat{a}_{k,\lambda}+{\bf E}_{-k}^{\lambda}(x,y)e^{-ikz}\hat{a}_{-k,\lambda}+{\rm h.c.}\big],\\
\hspace{-0.6cm}&&{\bf H}({\bf r})=\sum_{\lambda,k}\big[{\bf H}_k^{\lambda}(x,y)e^{ikz}\hat{a}_{k,\lambda}+{\bf H}_{-k}^{\lambda}(x,y)e^{-ikz}\hat{a}_{-k,\lambda}+{\rm h.c.}\big],
\end{eqnarray}
where the eigenmodes of electric and magnetic fields are expressed as \cite{electromagnetism}
\begin{eqnarray}
\hspace{-0.75cm}&&{\bf E}_k^{\lambda}(x,y,z)=\frac{1}{\sqrt{L}}\big[{\bgreek {\cal E}}_k^{t\lambda}(x,y)+{\bf {\cal E}}_{kz}^{\lambda}(x,y)\hat{\bf z}\big]e^{ikz},~~{\bf H}_{k}^{\lambda}(x,y,z)=\frac{1}{\sqrt{L}}\big[{\bgreek {\cal H}}_k^{t\lambda}(x,y)+{\bf {\cal H}}_{kz}^{\lambda}(x,y)\hat{\bf z}\big]e^{ikz},\\
\hspace{-0.75cm}&&{\bf E}_{-k}^{\lambda}(x,y,z)=\frac{1}{\sqrt{L}}\big[{\bgreek {\cal E}}_k^{t\lambda}(x,y)-{\bf {\cal E}}_{kz}^{\lambda}(x,y)\hat{\bf z}\big]e^{-ikz},~~{\bf H}_{-k}^{\lambda}(x,y,z)=\frac{1}{\sqrt{L}}\big[-{\bgreek {\cal H}}_k^{t\lambda}(x,y)+{\bf {\cal H}}_{kz}^{\lambda}(x,y)\hat{\bf z}\big]e^{-ikz}.
\end{eqnarray}
Here, ``$t$" means ``transverse", and $L$ is the length of the waveguide. The eigenmodes satisfy the following orthonormal relations \cite{electromagnetism},
\begin{eqnarray}
\nonumber
&&\int {\bgreek {\cal E}}_{k}^{t\lambda}\cdot {\bgreek {\cal E}}_{k'}^{t\mu}da=\delta_{kk'}\delta_{\lambda\mu}A_k^{\lambda},\\
\nonumber
&&\int {\bgreek {\cal H}}_{k}^{t\lambda}\cdot {\bgreek {\cal H}}_{k'}^{t\mu}da=\frac{1}{(Z_k^{\lambda})^2}\delta_{kk'}\delta_{\lambda\mu}A_k^{\lambda},\\
\nonumber
&&\int {\cal E}_{k,z}^{\lambda}{\cal E}_{k',z}^{\mu}da=-\frac{\gamma_{\lambda}^2}{k^2}\delta_{kk'}\delta_{\lambda\mu}A_k^{\lambda},\hspace{2.3cm}(\rm TM)\\
\nonumber
&&\int {\cal H}_{k,z}^{\lambda}{\cal H}_{k',z}^{\mu}da=-\frac{\gamma_{\lambda}^2}{k^2(Z_k^{\lambda})^2}\delta_{kk'}\delta_{\lambda\mu}A_k^{\lambda},\hspace{1.3cm}(\rm TE)\\
&&\frac{1}{2}\int ({\bgreek {\cal E}}_k^{t\lambda}\times {\bgreek {\cal  H}}_{k'}^{t\mu})\cdot\hat{\bf z}da=\frac{1}{2Z_k^{\lambda}}\delta_{kk'}\delta_{\lambda\mu}A_k^{\lambda},
\label{orthogonal}
\end{eqnarray}
where $Z_k^{\lambda}=\mu_0\omega_k^{\lambda}/k$  and $k/(\epsilon_0\omega_k^{\lambda})$ are the impedances for the TE and TM modes, $A_k^{\lambda}=\hbar \omega_k^{\lambda}/(2\epsilon_0)$ and $\hbar/(2\epsilon_0\omega_k^{\lambda})$ for the TE and TM modes with $\omega_k^{\lambda}$ being the eigen-energy, and 
\begin{equation}
\gamma_{\lambda}^2=\mu_0\epsilon_0(\omega_k^{\lambda})^2-k^2=(\omega_k^{\lambda})^2/c^2-k^2.
\label{spectra}
\end{equation} 
Please note that $\omega_k^{\lambda}=c\sqrt{k^2+\gamma_{\lambda}^2}$ in which $\gamma_{\lambda}$ only depends on the band but not the momentum is understood. With these orthonormal relations, we demonstrate 
\begin{equation}
\frac{\epsilon_0}{2}\int d{\bf r}{\bf E}({\bf r})\cdot{\bf E}({\bf r})+\frac{\mu_0}{2}\int d{\bf r}{\bf H}({\bf r})\cdot{\bf H}({\bf r})=\sum_{k,\lambda}\hbar\omega_{k}^{\lambda}(\hat{a}_{k,\lambda}^{\dagger}\hat{a}_{k,\lambda}
+\hat{a}_{-k,\lambda}^{\dagger}\hat{a}_{-k,\lambda}).
\end{equation}

With the established magnon and photon operators, the coupling Hamiltonian is constructed, 
\begin{equation}
\hat{H}_{\rm int}={\mu_0}\int d{\bf r}{\bf H}({\bf r})\cdot{\bf M}({\bf r})=
-{\mu_0\gamma\hbar}\int d{\bf r}(\hat{H}_x\hat{S}_x+\hat{H}_z\hat{S}_z).
\end{equation}
By assuming the magnetic sphere is small located at $(x_0,y_0,z_0)$, this Hamiltonian is calculated to be
\begin{equation}
\hat{H}_{\rm int}=\hbar \sum_{k_z=\pm k,\lambda}\big(g_{k_z,\lambda}\hat{a}_{k_z,\lambda}\hat{\alpha}^{\dagger}_K
+h_{k_z,\lambda}\hat{a}_{k_z,\lambda}\hat{\alpha}_K+{\rm h.c.}\big),
\end{equation}
where the coupling strength reads 
\begin{equation}
g_{k_z,\lambda}=\frac{\mu_0}{2}\sqrt{\frac{2\gamma M_sV_s}{\hbar L A_k^{\lambda}}}\big[i{\cal H}^{\lambda}_{k_z,x}(x_0,y_0)-{\cal H}^{\lambda}_{k_z,z}(x_0,y_0)\big]e^{ik_zz_0}.
\end{equation}

as we can see here, as the static field $H$ is assumed at y direction, the coupling effect between magnon and cavity photon states is determined by a combination of eigenmodes that perpendicular to $H$.

\subsection{Spontaneous radiation}

From above calculation, the whole Hamiltonian for a magnetic sphere in the waveguide is written as
\begin{equation}
H_0/\hbar=\omega_K \hat{\alpha}^{\dagger}_K\hat{\alpha}_K+\sum_{k_z}\omega_{k_z}\hat{a}_{k_z}^{\dagger}\hat{a}_{k_z}+\sum_{k_z}g_{k_z}(\hat{\alpha}_K^{\dagger}\hat{a}_{k_z}
+\hat{\alpha}_K\hat{a}_{k_z}^{\dagger}),
\end{equation}
in which $k_z=\pm k$ and the band index for photon is disregarded when only the lowest band is considered here. This Hamiltonian is known as the Fano-Anderson Hamiltonian with an exact solution being possible \cite{Fano,Mahan}. The lifetime of magnon can be calculated from the imaginary part of the self-energy \cite{Mahan} that is interpreted to come from the spontaneous radiation of magnon \cite{Fano}.
The Green function of Kittel magnon is exactly calculated to be
\begin{equation}
G_m(\omega)=\Big\{\hbar\omega-\hbar\omega_K+i\delta_K-\sum_{k_z}\frac{|\hbar g_{k_z}|^2}{\hbar\omega-\hbar\omega_{k_z}+i\delta_{k_z}}\Big\}^{-1},
\end{equation} 
where $\delta_K=\alpha\omega_K$ is the intrinsic Gilbert damping of Kittel magnon with $\alpha$ being the intrinsic Gilbert damping coefficient. 
The imaginary part of the self-energy, i.e., $\sum_{k_z}\frac{|g_{k_z}|^2}{\hbar\omega-\hbar\omega_{k_z}+i\delta_{k_z}}$, contributes to the broadening of the magnon spectra. Therefore the total broadening of magnon reads
\begin{equation}
\Delta E_K=\delta_K+\pi \sum_{k_z}\big|\hbar g_{k_z}\big|^2\delta(\hbar\omega_K-\hbar\omega_{k_z})=
\delta_K+\frac{\pi}{\hbar}\big|\hbar g(\omega_K)\big|^2{\cal D}(\omega_K),
\end{equation}
where ${\cal D}(\omega_K)$ is the global density of states (DOS) of photon in the waveguide.

By further defining the width in terms of the magnetic field $\Delta E=\hbar\gamma\mu_0\Delta H$, we obtain
\begin{equation}
\mu_0\Delta H=\frac{\alpha \omega_K}{\gamma}+\frac{\pi}{\gamma}|g(\omega_K)|^2{\cal D}(\omega_K)=\frac{\alpha \omega_K}{\gamma}+\frac{2\pi}{\gamma}\xi|g_{\rm CST}(\omega_K)|^2.
\label{final}
\end{equation}
For TE mode in the experiment, parameter $\xi$ is expressed as 
\begin{equation}
\xi=\frac{1}{2}\Big(\frac{\mu_0}{2}\sqrt{\frac{2\gamma M_s V_s}{\hbar}}\sqrt{\frac{\hbar \omega_K}{2\epsilon_0}}\frac{1}{c\sqrt{\pi\mu_0}}\Big)^2.
\end{equation}
In Eq.~(\ref{final}), another coupling strength $g_{\rm CST}$ is defined when the normalization with unit power flow in waveguide is used in the {\it Computer simulation technology} (CST) \cite{electromagnetism}, i.e.,
\begin{equation}
\frac{1}{2}\int (\tilde{{\bgreek {\cal E}}}_k^{t\lambda}\times \tilde{{\bgreek {\cal  H}}}_{k'}^{t\mu})\cdot\hat{\bf z}da=\frac{1}{2}\delta_{kk'}\delta_{\lambda\mu}.
\end{equation}
The relation between $\{{\cal \tilde{E}}_k,{\cal \tilde{H}}_k\}$ here and $\{{\cal E}_k,{\cal H}_k\}$ in Eqs.~(\ref{orthogonal}) reads
\begin{equation}
\{{\cal \tilde{E}}_k,{\cal \tilde{H}}_k\}=\{{\cal E}_k,{\cal H}_k\}\times c\sqrt{\pi \mu_0}\sqrt{{\cal D}/(A_kL)}.
\end{equation}
It is seen that $g_{\rm CST}$ is proportional to the local field strength and $\sqrt{\cal D}$.

\subsection{Relation to local density of states}

Although the CST does not directly calculate the local density of states (LDOS), we can show $|g_{\rm CST}|^2$ is indeed proportional to the LDOS. To define the magnetic-field LDOS {\it in the TE mode} \cite{LDOS}, we define the new eigenmodes with different orthonormal conditions from Eqs.~(\ref{orthogonal}), 
\begin{eqnarray}
\nonumber
&&\int (\overline{{\bgreek {\cal H}}}_{k}^{t\lambda})^*\cdot \overline{{\bgreek {\cal H}}}_{k'}^{t\mu}da=\int \overline{{\bgreek {\cal H}}}_{k}^{t\lambda}\cdot \overline{{\bgreek {\cal H}}}_{k'}^{t\mu}da=\delta_{kk'}\delta_{\lambda\mu}(B_k^{\lambda})^2/{\mu_0},\\
&&\int (\overline{{\cal H}}_{k,z}^{\lambda})^*\overline{{\cal H}}_{k',z}^{\mu}da=-\int \overline{{\cal H}}_{k,z}^{\lambda}\overline{{\cal H}}_{k',z}^{\mu}da=\frac{\gamma_{\lambda}^2}{k^2}\delta_{kk'}\delta_{\lambda\mu}(B_k^{\lambda})^2/{\mu_0},
\label{orthogonal2}
\end{eqnarray}
where $B_k^{\lambda}=ck/\omega_{k}^{\lambda}$. 
This means 
\begin{equation}
\overline{\cal H}_{k}^{\lambda}={\cal H}_k^{\lambda}\times \frac{Z_k^{\lambda}B_k^{\lambda}}{\sqrt{\mu}_0\sqrt{A_k^{\lambda}}}.
\end{equation}
With these conditions, for the waveguide with uniform section, i.e., with translation symmetry,
the magnetic-field LDOS per unit length is defined by \cite{LDOS}
\begin{equation}
\rho_{\alpha}(x,y,\omega)=\frac{1}{L}\mu_0\sum_{k,\lambda}\delta(\omega-\omega_{k}^{\lambda})|\overline{{\cal H}}_{k,\alpha}^{\lambda}(x,y)|^2,
\end{equation}
with $\alpha=\{x,y,z\}$.
From the normalization condition in Eqs.~(\ref{orthogonal2}), it is checked that $\sum_{\alpha}\int dxdy \rho_{\alpha}(x,y,\omega)={\cal D}(\omega)/L$. When $\lambda$ is the lowest TE mode, we find that
\begin{equation}
\rho_{\alpha}(x,y,\omega)=\frac{1}{L}\mu_0{\cal D}(\omega)|\overline{{\cal H}}_{k,\alpha}(x,y)|^2.
\end{equation}
This leads to
\begin{eqnarray}
\nonumber
&&|g(\omega)|^2{\cal D}(\omega)=\xi|g_{\rm CST}(\omega)|^2=\Big(\frac{\mu_0}{2}\Big)^2\frac{2\gamma M_sV_s}{\hbar LA_k^{\lambda}}\big[|{\cal H}_z^{k}(x_0,y_0)|^2+|{\cal H}_x^{k}(x_0,y_0)|^2\big]{\cal D}(\omega)\\
&&=\frac{\gamma M_sV_s}{2\hbar c^2}\rho_\perp(x_0,y_0,\omega)\equiv \kappa \rho_\perp(x_0,y_0,\omega),
\end{eqnarray}
where $\kappa=\frac{\gamma M_sV_s}{2\hbar c^2}$ and $\rho_\perp(x_0,y_0,\omega)=\rho_z(x_0,y_0,\omega)+\rho_x(x_0,y_0,\omega)$. Thus, for waveguide with uniform section,
\begin{equation}
\rho_\perp(x_0,y_0,\omega)=\xi|g_{\rm CST}|^2/\kappa,
\end{equation}
that can be determined by CST simulation approach. Finally, we can reach
\begin{equation}
\mu_0\Delta H=\mu_0\Delta H_0+\frac{\alpha \omega_K}{\gamma}+\frac{2\pi \kappa}{\gamma}R|\rho_\perp(x_0,y_0,\omega)||\mathcal{LDOS}|,
\label{broadening_LDOS}
\end{equation}
in which $|\mathcal{LDOS}|\equiv 1$~$\rm m^{-3}\cdot s$ explicitly represents the unit of LDOS for photons, $\mu_0\Delta H_0$ is the inhomogeneous broadening of magnon linewidth at zero frequency and $R$ is a fitting parameter mainly influenced by cavity design and cable loss in practical measurement circuit.  

Above theory is established for the ideal waveguide with the same cross section along the waveguide. When the cross section slowly varies as the case in our experiment, this treatment can be still a good approximation, with the fitting parameter quantity $R\sim0.8$ that is close to unity. A direct comparison between the theory and experiment in ideal waveguide is presented in the following section.


\section{Supplementary Note 2.\\
	Magnon linewidth enhancement in rectangular waveguide}

\subsection{Enhanced DOS near waveguide cut-off frequency}

In this part, the enhancement of DOS of photon near the cut-off frequency of a standard rectangular waveguide is measured.  A rectangular waveguide that works at Ku band ($a=15.8$~mm and $b=7.8$~mm) is used for constructing photon states, as schematically shown in Fig. \ref{Fig1}. The transmission $|S_{21}|$  as well as the reflection $|S_{11}|$ signals are measured by an Vector Network analyzer (VNA) with an input power of 0~dbm, as shown in  Fig. \ref{Fig1}(a). $S_{21}$ amplitude is observed to decrease below the cut-off frequency $\omega_{\rm cut}/2\pi$=9.8~GHz of our device because the microwave becomes evanescent with larger wavelength.

\begin{figure*}  [!htbp]
	\begin{center}
		\epsfig{file=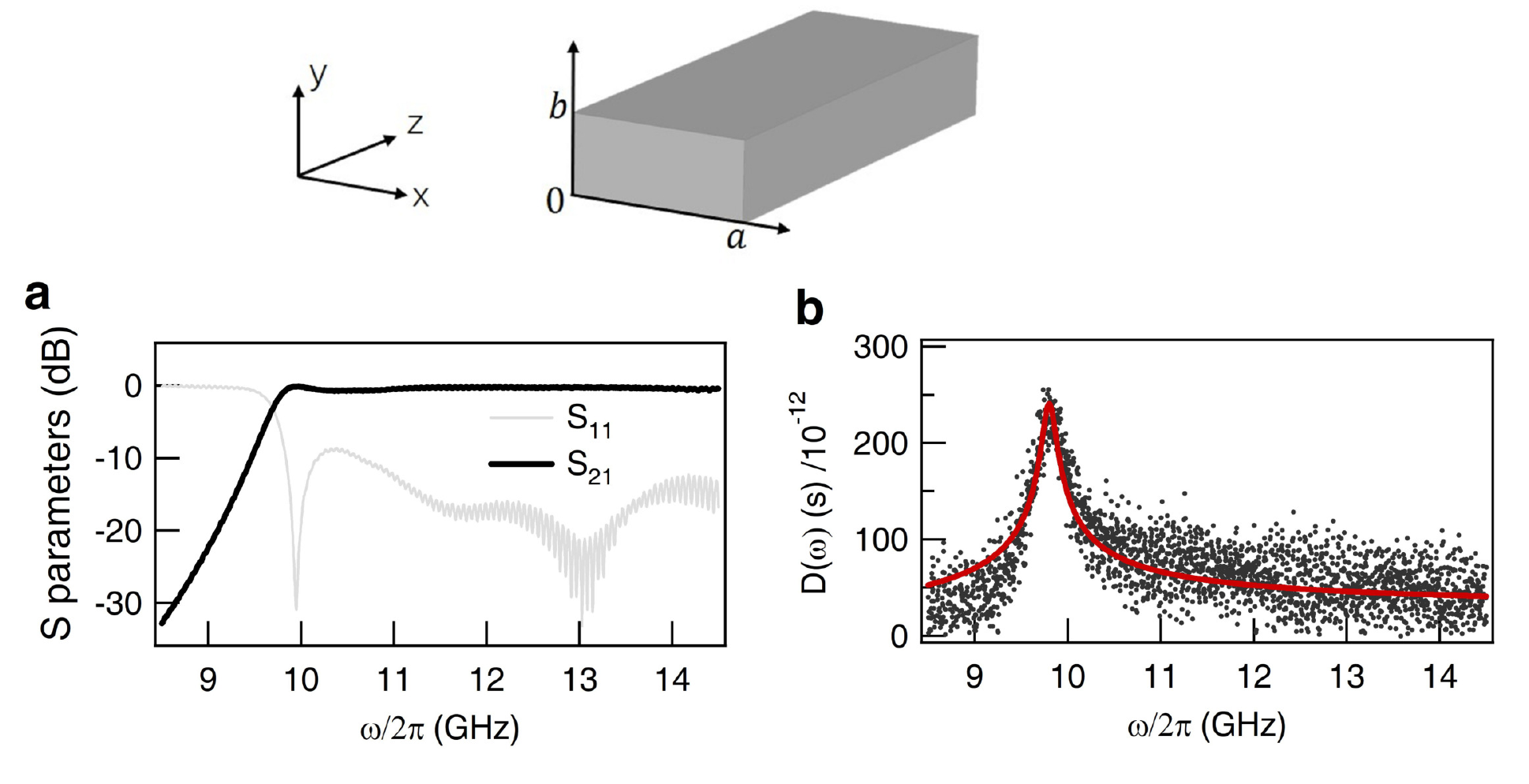,width=15cm}
		\caption{(Color online) (a) Measured $S$-matrix parameters for Ku-band rectangular waveguide. (b) Caculated microwave photon DOS for the rectangular waveguide (dots) with fit line (solid lines).}\label{Fig1}	\end{center}
\end{figure*}

Physically, near the cut-off frequency, the magnitude of the momentum for the microwaves tends to be zero along the propagation direction, leading to an enhancement of global DOS $D(\omega)$ near cut-off frequency. This effect can be verified by obtaining the global DOS from $S$-matrix parameters via the inversion relation \cite{Genack2015,Genack1999},
\begin{equation}
{\cal D}(\omega)=-i(1/2\pi V){\rm Tr}(S^\dagger dS/d\omega),
\label{FromS}
\end{equation}
where $S(\omega)=\begin{pmatrix} S_{11}(\omega) &S_{12}(\omega) \\S_{21}(\omega) & S_{22}(\omega) \end{pmatrix}$ represents the scattering matrix. From the measured $S$ parameters, we can plot DOS in Fig. \ref{Fig1}(b), in which an enhancement of DOS near the cut-off frequency is clearly observed. 

Analytically, the global DOS for photons in the ${\rm TE}_{10}$ mode in a rectangular
waveguide is calculated to be \cite{electromagnetism}
\begin{equation}
{\cal D}(\omega)=\sum_{k_z}\delta(\omega-\omega_{k_z}^{(1,0)})
=\sum_{k_z}\delta(\omega-c\sqrt{(\pi/a)^2+k_z^2})=\frac{L}{\pi}\frac{\omega}{c\sqrt{\omega^2-(c\pi/a)^2}}.
\end{equation}
For further considering the energy loss caused by the finite conductivity in the waveguide wall, the DOS spectrum can be well fit by taking consideration of the damping factor $\beta$. As shown in Fig.~\ref{Fig1}(b) , the DOS spectrum is well reproduced by $D(\omega)=\frac{L}{\pi}\frac{\omega}{c\sqrt{\omega^2-\omega_{\rm cut}^2+i\beta\omega\omega_{\rm cut}}}$, with the damping $\beta$ and propagation distance $L$ fitted by $0.0159$ and $0.03$~m, respectively.

\subsection{Magnon linewidth enhancement near mode cut-off}

In this part, the magnon linewidth enhancement near waveguide cutoff frequency can be directly tested in our standard rectangular waveguide, in which the section is uniform along the waveguide. This allows a direct comparison between the theory [Eq.~(\ref{broadening_LDOS})] and experimental observation without the approximation that the section varies slowly along  the waveguide.

We introduce a 1~mm highly-polished YIG sphere into the rectangular waveguide to display the damping control via photon states. We place the YIG sphere in two different positions, i.e., at the cross-section center and close to the conductive wall of the waveguide. By fitting the linewidth of $|S_{21}(H)|^2$ spectra at different microwave frequencies $\omega$, we can plot the magnon linewidth $\mu_0\Delta H$ as a function of $\omega$ in Fig. \ref{Fig2} for two different positions, respectively.

\begin{figure*}  [!htbp]
	\begin{center}
		\epsfig{file=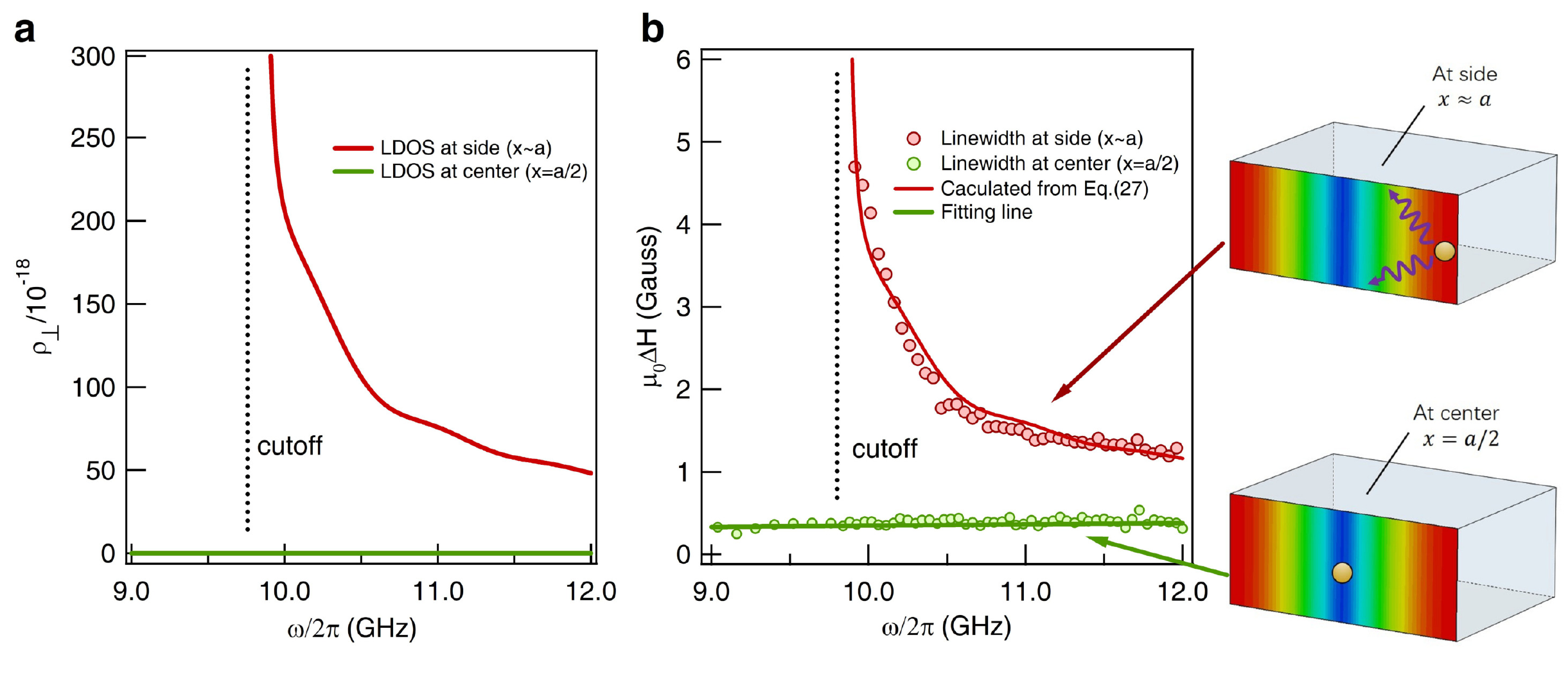,width=15cm}
		\caption{(Color online) (a) Simulated LDOS $\rho_\perp$ when placing YIG sphere at waveguide center and near waveguide wall, respectively. (b) Measured magnon linewidth for two different positions with the YIG sphere placed in the center $x=a/2$ (green dots) and close to conductive wall $x\sim a$ (red dots).  Red solid curve in (b) is calculated from Eq.~(\ref{broadening_LDOS}) in supplementary note~1 and the green solid curve is fitted for intrinsic linewidth.}\label{Fig2}
	\end{center}
	
\end{figure*}

At the cross-section center with $x=a/2$, the microwave LDOS is found to be nearly zero [see the green curve in Fig. \ref{Fig2}(a)] and therefore LDOS-induced radiation damping can be approximately neglected here. As a result, intrinsic magnon linewidth can be obtained from our measurement, with a typical linear $\Delta H-\omega$ revealed in Fig. \ref{Fig2}(b) by the green circles. Using linear fitting, the Gilbert damping coefficient can be obtained as 4.3$\times10^{-5}$ with the inhomogeneous broadening $\mu_0\Delta H_0$ fitted by 0.19~Gauss. 

As shown in Fig.~\ref{Fig2}(a) by the red curve, LDOS near waveguide wall displays larger magnitude with an obvious enhancement near the mode cut-off frequency. Hence by moving the YIG sphere to near the conductive wall, we expect and indeed obtain a magnon linewidth enhancement according to Eq.~(\ref{broadening_LDOS}). 
Specifically, we plot the measured magnon linewidth as a function of frequency in Fig. \ref{Fig2} (b) by the red circles.  Compared to $\Delta H$-$\omega$ relation when YIG is placed at center, here we observe significant  magnon linewidth enhancement especially at the frequency near the mode cut-off. Such linewidth broadening can be well reproduced by our calculation from Eq.~(\ref{broadening_LDOS}) with the fitting parameter $R$ close to unity.

\end{widetext}

\end{document}